\documentclass{aa}
\usepackage{graphicx, natbib}
\bibpunct[]{(}{)}{;}{a}{}{,}

\begin{document}

\title{The relationship between star formation rates \\
       and mid-infrared emission in galactic disks\thanks{Based on observations
       with ISO, an ESA project with instruments funded by ESA Member States
       (especially the PI countries: France, Germany, the Netherlands
       and the United Kingdom) and with the participation of ISAS and NASA.}}

\author{H. Roussel\inst{1}
\and M. Sauvage\inst{1}
\and L. Vigroux\inst{1}
\and A. Bosma\inst{2}}

\institute{DAPNIA/Service d'Astrophysique, CEA/Saclay, 91191 Gif-sur-Yvette cedex, France
\and Observatoire de Marseille, 2 Place Le Verrier, 13248 Marseille cedex 4, France}

\authorrunning{\hspace*{-0.5cm} H. Roussel et al.}
\titlerunning{Relationship between SFRs and MIR emission in galactic disks}

\offprints{H. Roussel (e-mail: hroussel@cea.fr)}

\date{Received 2 February 1999 / Accepted 19 March 2001}

\maketitle

\begin{abstract}
The H$\alpha$ and mid-infrared mean disk surface brightnesses are compared in a
sample of nearby spirals observed by ISOCAM. This shows that, in spiral disks,
dust emission at 7 and 15\,$\mu$m provides a reasonable star formation tracer.
The fact that the 15 to 7\,$\mu$m flux ratio is nearly constant in various
global exciting conditions indicates a common origin, namely the aromatic
infrared band carriers, and implies that at these wavelengths, dust emission
from the disks of normal galaxies is dominated by photodissociation regions
and not by H{\scriptsize II} regions themselves. \\
We use this newly-found correlation between the mid-infrared and the H$\alpha$
line to investigate the nature of the link between the far-infrared (60 and
100\,$\mu$m) and H$\alpha$. Although the separation of the central regions from the
disk is impossible to achieve in the far-infrared, we show that a circumnuclear
contribution to the dust emission, having no equivalent counterpart in H$\alpha$,
is most likely responsible for the well-known non-linearity between far-infrared
and H$\alpha$ fluxes in spiral galaxies. \\
We derive a calibration of 7 and 15\,$\mu$m fluxes in terms of star formation
rates from a primary calibration of H$\alpha$ in the literature, and also outline
the applicability limits of the proposed conversion, which should not be
blindly extrapolated to objects whose nature is unknown.
\end{abstract}

\section{Introduction}
\label{sec:intro}

Whether mid-infrared emission can be considered a reliable tracer of
the massive stellar content of normal and isolated spirals is still
unclear. The accepted interpretation of mid-IR spectra of galaxies
\citep[see the review by][]{Puget, Desert} is that they
consist primarily of a composite of a featureless continuum and of a
family of aromatic bands, the so-called unidentified infrared bands
(UIBs). The continuum emission is attributed to very small grains, VSGs
\citep{Desert}, of which little is known,
while various carbonaceous materials have been suggested as candidates
to carry the UIBs, among which the PAH model (polycyclic aromatic
hydrocarbons) of \citet{Leger} has been a long-time favorite.
However, the recent work of \citet{Boulanger2} indicates that UIB
carriers are likely aggregates of several hundred atoms rather than
macro-molecules. It is important to realize that under most radiation
field conditions, both components are out of thermal equilibrium and
undergo large temperature fluctuations of several hundred K.

\citet{Sturm} have provided a census of the continuum emission and of
emission features (UIBs and ionic lines) found from 2.4 to 45\,$\mu$m in
typical starburst galaxies, and which are potentially present in our data
as well. Emission from the envelopes of cold stars can also contribute
in the 7\,$\mu$m bandpass, but it is negligible in our sample, except
possibly in the disks of two S0/a--Sa galaxies.

This dual nature of the mid-infrared emission (produced mainly by two
dust phases, UIB carriers and VSGs) makes the existence of a direct
link with massive stars unlikely. A further complication is that even
if both species are predominantly heated by high energy radiation,
their excitation by optical and near ultraviolet photons may be
significant in environments where old stellar populations dominate.
Indeed, aromatic bands are ubiquitous in the diffuse interstellar
medium \citep{Giard, Mattila} and are also observed in
regions where the ultraviolet radiation density is insufficient to
account for their heating \citep{Sellgren, Boulade, Uchidaa, Uchidab}.
When observing extragalactic
objects, emission arising in star forming regions is mixed with
that arising in the interstellar medium associated with more evolved
stars. Therefore, the accuracy with which mid-infrared emission
traces star formation should in principle depend on the balance
between these two heating sources. This can be checked either by
investigating spatially resolved galaxies, or by building global
energy budgets. As we will show later, the flux fraction due to
heating by old stellar populations is generally quite small in spiral
galaxies, as already noted by \citet{Lemke}.

An attempt to link the mid-infrared emission of galaxies with recent
star formation is, however, encouraged by the following fact: resolved
observations of individual regions in our Galaxy have revealed
that aromatic bands are closely associated with the photodissociation
shells and surfaces of molecular clouds in the vicinity of H{\scriptsize II}
regions or hot stars, while the VSG continuum strongly peaks inside
H{\scriptsize II} regions \citep{Cesarsky, Tran, Verstraete}.
Given that both types of sites are intimately linked with the presence
of massive stars, a strong coupling between mid-infrared emission and
present-day star formation could exist.

An invaluable advantage of infrared observations over optical
recombination lines or the ultraviolet continuum resides in their much
lower sensitivity to interstellar extinction, thus providing insights
into obscured star forming regions. Besides, if it is confirmed that
young stars are the major heating source of dust emitting at 7 and
15\,$\mu$m, mid-infrared fluxes could provide a more acceptable star
formation tracer than far-infrared fluxes, since the latter have been
shown to contain a cirrus component \citep{Helou86} which dominates
the mean emission from morphological types Sa to Sc
\citep{Sauvage92} and is responsible for a strong non-linearity in
the correlation between far-infrared and H$\alpha$ fluxes. A vast
literature covers the advantages and limitations of interpreting
far-infrared emission as a star formation indicator, and includes for
instance \citet{Lonsdale}, \citet{Devereux} and \citet{Smith96}. A
summary of the issues in question can be found {\it e.g.} in
\citet{Sauvage92} or in \citet{Kennicutt98a}.

On the other hand, since survey programs performed with ISO have
focussed on observations at 7 and 15\,$\mu$m, investigating the
relationship between the emission in these bandpasses and star
formation rates would be very helpful for their physical
interpretation.

To tackle this question, we use a sample of 69 galaxies observed by ISOCAM
in two broadband filters centered at 7 and 15\,$\mu$m. All the galaxies being
nearby, the achieved spatial resolution is sufficient to delineate distinct
structural entities (spiral arms, giant H{\scriptsize II} complexes, the
circumnuclear concentration, etc.). Distances to the sample galaxies range from
4 to 60\,Mpc, which translates into linear resolutions of 110\,pc to 1.7\,kpc
at 7\,$\mu$m (full width at half maximum of the point spread function). This allows
a clear differentiation of the mid-infrared properties of central regions and
disks, a study described in Roussel et al. (2000a and b, hereafter Atlas and
Paper~I). In this paper, we use this advantage to restrict ourselves to the
study of galactic disks. Our motivation for that is manifold: \\
(1) The mid-infrared colors of galaxies in our sample are relatively uniform in
the disks, while the circumnuclear regions can show strong 15\,$\mu$m emission
excesses, suggesting the existence of a different thermodynamical state of
dust in the central parts of galaxies. \\
(2) Since we want to establish a calibration of mid-infrared fluxes in terms of
star formation rates, we need to compare them with a direct primary star
formation tracer and so far, few of them sample equally well the disk and
the nuclear regions of galaxies, mainly because of extinction. \\
(3) It has been shown \citep{Kennicutt98a} that star formation processes
and physical conditions prevailing in nuclei are widely different from those
of disks.

In Sect.~\ref{sec:sample} we present our sample and the methods used to
collect the photometric information needed for our analysis. Sect.~\ref{sec:sfr}
demonstrates the validity of mid-infrared fluxes as star formation tracers
in galactic disks. In Sect.~\ref{sec:dis} we discuss the applicability limits
of our calibration and the implications of our findings on the
interpretation of the far-infrared emission of galaxies.

\section{The galaxy sample and photometric data}
\label{sec:sample}

   The sample of spiral galaxies considered here is made from the partial
merging of five ISOCAM programs: \\
-- {\it Camspir} (PI: L. Vigroux) which mapped nearby very large spiral
   galaxies extensively observed in other interstellar tracers, allowing
   detailed spatial analyses. \\
-- {\it Cambarre} (PI: C. Bonoli) which mapped barred spiral galaxies, selected
   to span the variety of bar and Hubble types and a large range of infrared
   luminosities. \\
-- resolved spirals from the complete survey of the {\it Virgo} program
   \citep[PI: J. Lequeux; see][]{Boselli}. \\
-- non-Seyfert spirals, with no strong signs of tidal interaction, from
   the {\it Sf\_glx} program \citep[PI: G. Helou; see][]{Dale} which selected
   galaxies sampling the IRAS color-color diagram \citep{Helou86}. \\
-- identically chosen spirals from the {\it Galir} program (PI: T. Onaka) that
   aimed at constructing infrared spectral energy distributions of normal
   galaxies, in preparation for the Japanese mission IRTS.

The resulting sample comprises 69 galaxies spanning the whole de Vaucouleurs
spiral sequence from S0/a to Sdm. All were observed in raster mode in the two
filters LW2 centered at 7\,$\mu$m (5--8.5\,$\mu$m) and LW3 centered at 15\,$\mu$m
(12--18\,$\mu$m), with a pixel size of $3\arcsec$ or $6\arcsec$. General information
and a deeper discussion of the mid-infrared properties of these galaxies can be
found in Paper~I, together with the spectra between 5 and 16\,$\mu$m observed in
five of them. The detailed description of data reduction appears in the Atlas,
that also presents the 7\,$\mu$m maps. We simply note here that we processed all
maps in a homogeneous way, including those already published.

\subsection{H$\alpha$ data}
\label{sec:ha}

Since the goal of this paper is to assess the reliability of mid-infrared
emission as a star formation tracer, we need to collect data on a primary
indicator. Recombination lines, which trace the existence of massive stars,
are an obvious choice: well-established calibrations in terms of star formation
rate (SFR) exist, and their production by post-AGB stars, as seen in
ellipticals \citep{Binette}, is negligible in star-forming galaxies
\citep{Kennicutt98a}.

We therefore searched the literature for integrated H$\alpha$ photometry.
In addition, some H$\alpha$ maps were kindly made available to us by J.A.
Garc\'{\i}a-Barreto for NGC\,1022 and NGC\,4691 \citep[published in][]{Garcia};
T. Storchi-Bergmann for NGC\,1097 and NGC\,1672 \citep[published in][]{Storchi};
M. Naslund for NGC\,1365 \citep[published in][, courtesy of S. J\"ors\"ater,
M. Naslund and J.J. Hester]{Lindblad}; C. Feinstein for
NGC\,7552 \citep[data published in][]{Feinstein}; M.W. Regan via D. Reynaud
for NGC\,1530 \citep[published in][]{Regan}; S.D. Ryder via A. Vogler for
NGC\,5236 \citep[published in][]{Ryderb, Vogler}; F. Viallefond for NGC\,5457.
Some of these maps have been
corrected for a spatial gradient, using the background, or for an
over-subtraction of the continuum emission, using I-band images. In some
cases, we also performed the flux-calibration, using data from the
literature inside various apertures.

Concerning galaxies for which we have no map, the bulk of the data comes from
\citet{Young}. We note that this reference provides (H$\alpha$~+~[N{\scriptsize II}]) fluxes
systematically higher than those of \citet{Kennicutt83b} for the galaxies in
common (by a factor ranging from about one to two). We thus preferred to adopt
data from \citet{Young} or other references, but for six of our galaxies, they
were taken from \citet{Kennicutt83b}; we have corrected them by a factor 1.16,
following \citet{Kennicutt98b}'s prescription, and we have checked that the
optical diameter of these galaxies is less than or comparable to the H$\alpha$
aperture used (except for VCC\,2058: the optical size and H$\alpha$ aperture are
respectively $4.27\arcmin$ and $3\arcmin$).

Since the central regions of most galaxies in our sample stand out in the
mid-infrared as having different properties from the disk (see Paper~I and the
Atlas) and as the contamination of H$\alpha$ fluxes by nuclear regions can be
significant in galaxies harboring non-stellar activity or starburst, and for
all the reasons emphasized in Sect.~\ref{sec:intro}, we chose to exclude the
central regions from both mid-infrared and H$\alpha$ measurements. For this
purpose, we used matched apertures which were dictated by the available H$\alpha$
data in the literature. We aimed at subtracting circumnuclear fluxes of
sufficiently large a region to match that region inside which most of a possible
15\,$\mu$m excess is located. It was straightforward to achieve this when we
could directly perform measurements on H$\alpha$ maps. However, this could not
be achieved in practice for all galaxies, due to the difficulty of finding
suitable H$\alpha$ data. When such nuclear data are not available or were
measured only inside an aperture significantly smaller than the size of the
circumnuclear region in the mid-infrared (this concerns 8 galaxies out of 44),
we checked that the circumnuclear $F_{15}/F_7$ color is low (below 1.2),
{\it i.e.} shares the main characteristic of disks. This ensures that we are not
introducing a strong bias, because the color indicates that the star formation
process and extinction should be close to those found in disks. The aperture
used by \citet{Pogge} (from whom the central H$\alpha$ fluxes of 8 galaxies
are obtained) is not given explicitly, but could be estimated from his H$\alpha$
images, except for VCC\,460 and VCC\,857, for which we have assumed that it is
equal to the infrared size of the central region. This could be problematic in
the case of VCC\,460, whose central color is high ($F_{15}/F_7 = 2.55$). However,
all these uncertainties remain a negligible source of error with respect to the
extinction correction.

Most H$\alpha$ measurements (at 6563\,\AA) include the two neighboring
[N{\scriptsize II}] lines (the most intense at 6583\,\AA\ and another one at
6548\,\AA). We applied the same correction as \citet{Kennicutt83a} for average
[N{\scriptsize II}] contamination in disk H{\scriptsize II} regions (25\% of the
total flux). The [N{\scriptsize II}]/H$\alpha$ ratio is in general higher in
central regions than in disk H{\scriptsize II} regions \citep[{\it e.g.} ][]{Brand},
but we have removed central ([N{\scriptsize II}]~+~H$\alpha$) fluxes, and ratios
in disks are little dispersed \citep{Kennicutt83b}.

We also made use of the value given by \citet{Kennicutt83a} for average and
uniform extinction in the H$\alpha$ line (expected to be the major source of
uncertainty, since it amounts to 1.1\,mag). It is clear that a uniform extinction
correction is in principle very far from the true correction that should be
applied. However, we can first expect that regions where the extinction most
significantly departs from this value are located in the central parts of
galaxies, which we have excluded from the present analysis. Second, it is on
H$\alpha$ data corrected in this way that SFR calibrations are built. And third,
apart from observationally deriving the extinction in each object, it is not
possible to define a correction scheme ({\it e.g.} based on Hubble type, or on
global Balmer decrement) that does not introduce as much uncertainty and bias
as it supposedly removes. We therefore choose to confine ourselves to this
uniform correction scheme, bearing in mind that our conclusions are relative
to this method of correcting H$\alpha$ data in order to estimate the SFR.

Control on the bias that we introduce thereby can be found in an examination
of the H$\alpha$ to 15\,$\mu$m flux ratio (since we apply a uniform correction on
the H$\alpha$ data, it makes no difference here whether these are corrected or not).
The variation of the ratio of H$\alpha$ to 15\,$\mu$m fluxes as a function of the
inclination (estimated from kinematical data or, if unavailable, from the ratio
of major to minor isophotal diameters) is a pure scatter diagram. We also checked
that separating our sample into two morphological classes (22 S0/a--Sb and
20 Sbc--Sdm) did not result in significant a difference in the H$\alpha$ to
15\,$\mu$m flux ratio: the logarithmic means in the two subsamples are respectively
$-0.67 \pm 0.18$ and $-0.66 \pm 0.21$. Finally, no trend can be seen in a plot
of the H$\alpha$ to 15\,$\mu$m flux ratio versus the size-normalized
H$\alpha$ flux (see section~\ref{sec:sfr} for a definition of this quantity).
Since the 15\,$\mu$m emission is much less prone to absorption, we are therefore
confident that no systematic variation in the extinction affects the H$\alpha$ data.
Furthermore, the values of $F_{15} / F$(H$\alpha$) in our sample are all compatible
with moderate absorption if compared with the values observed in M\,51
\citep{Sauvage96}.

Galactic extinction was corrected using the blue absorptions listed in
the RC3 together with the extinction curve of \citet{Cardelli}. The galaxies
for which suitable H$\alpha$ data were found are listed in Table~\ref{tab:tableau}
with their total and central fluxes.

\subsection{Mid-IR photometry}

We have measured total fluxes as explained in the Atlas, and central fluxes
inside the same aperture as that used for H$\alpha$, in images treated with an
algorithm designed to correct for dilution effects, also described in the
Atlas. The disk flux is then the difference between the total flux and the
corrected nuclear flux. The resulting central flux fractions which have been
removed and disk fluxes are given in Table~\ref{tab:tableau}.

As already mentioned, the aperture of H$\alpha$ measurements matches
reasonably well the size of the circumnuclear concentration in the
images corrected for dilution. We call the attention of our readers to
the fact that the size of the aperture used to measure central mid-IR
fluxes in this paper is different from that used in Paper~I. In Paper~I,
since we did not have the constraint to match the mid-IR aperture to that
available for data taken at a different wavelength, the size of the
central regions was measured on non-deconvolved mid-IR surface brightness
profiles. The aperture sizes given in Table~\ref{tab:tableau} are thus
different from those used in Paper~I.

Typical errors on disk fluxes are expected to be of the order of 20 or
30\%, mainly due to the camera memory effects\footnote{The stabilization
correction that we applied uses the latest available technique taking into
account the detector characteristics, described in \citet{Coulais}.} and to
flux calibration uncertainties.

\section{The star formation rate scaling in galactic disks}
\label{sec:sfr}

In M\,51 and NGC\,7331 for example, there exists a striking correspondence
between the spatial distributions of H$\alpha$ and 15\,$\mu$m emission
\citep{Sauvage96, Smith98}. When we compare in detail our 7 and 15\,$\mu$m maps
with the H$\alpha$ maps at our disposal, all structures, such as rings,
bars and arms, are very similar at the three wavelengths, even without correction
for the different angular resolutions and sensitivities. Thus, it is tempting
to check the robustness of this correlation in a more quantitative approach.
If it holds for total fluxes, it would indicate a strong relationship between
mid-infrared dust emission and the SFR.

We first have to cancel identifiable bias sources in our data.
In addition to the varying contribution of the central regions,
the most obvious one is a scale effect, {\it i.e.} large -- or bright --
galaxies tend to be bright at all wavelengths, and comparing them with
small and faint galaxies introduces artificial correlation of the
data. Another bias results from the use of luminosities instead of
fluxes: an incorrect estimation of distances introduces dispersion and
the presence of the distance squared on both axes also produces an
artificial correlation. We therefore have to normalize our data by
another galaxy property, independent of both mid-infrared and H$\alpha$
emission. We chose for such a quantity the disk area (from the major
diameter at the blue isophote $\mu_{\rm B} = 25$\,mag\,arcsec$^{-2}$,
defined in the RC3). We stress that even though the normalization of
fluxes by the disk area gives quantities that are formally surface
brightnesses, these should not be identified with mid-infrared or H$\alpha$
surface brightnesses. The normalization is only used to avoid the
scale-effect in our sample. In the following, we will refer to the
quantities obtained in this way as ``size-normalized'' fluxes.

\begin{figure}[!t]
\centerline{\resizebox{9cm}{!}{\rotatebox{90}{\includegraphics{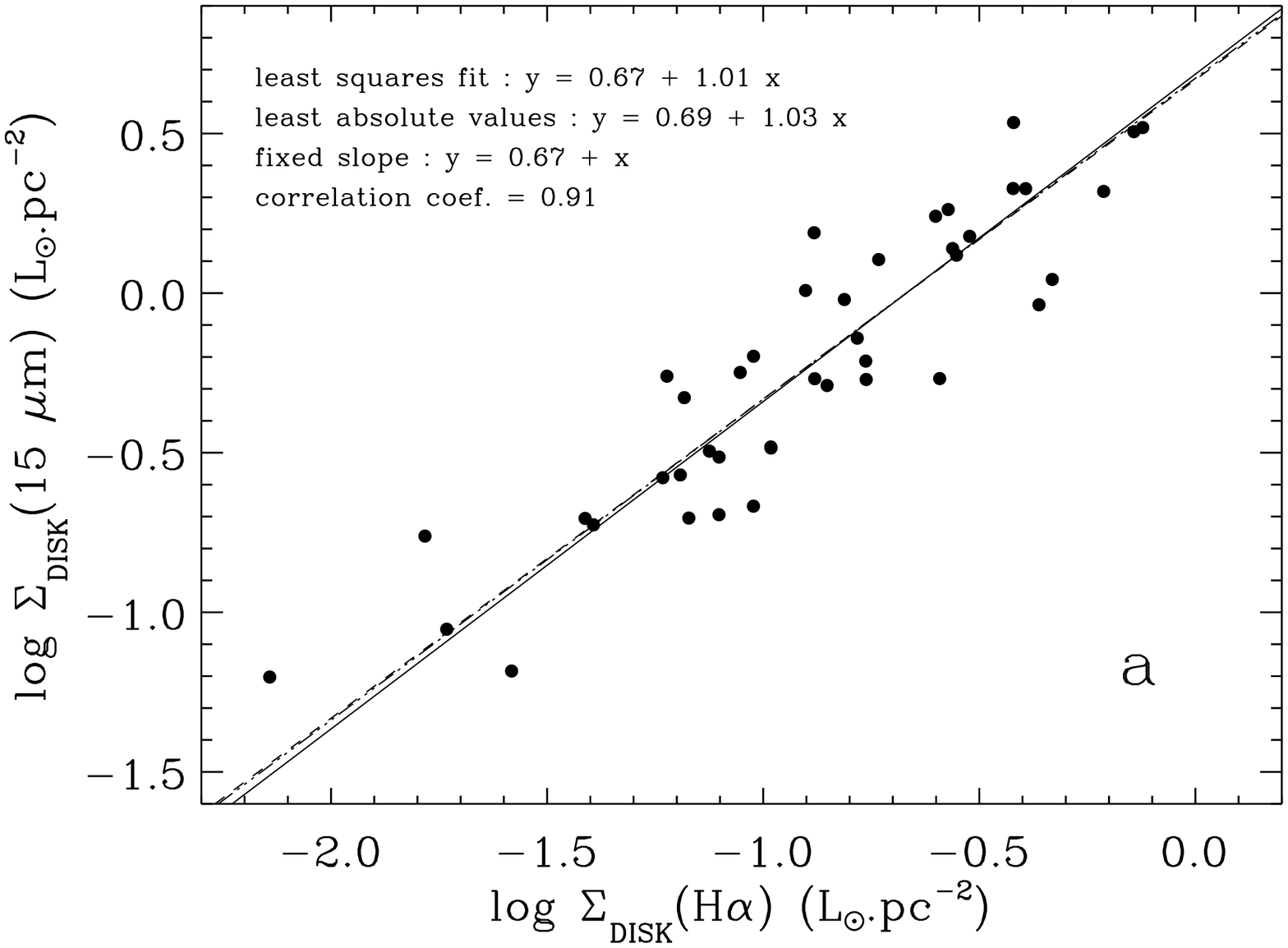}}}}
\vspace*{-0.5cm}
\centerline{\resizebox{9cm}{!}{\rotatebox{90}{\includegraphics{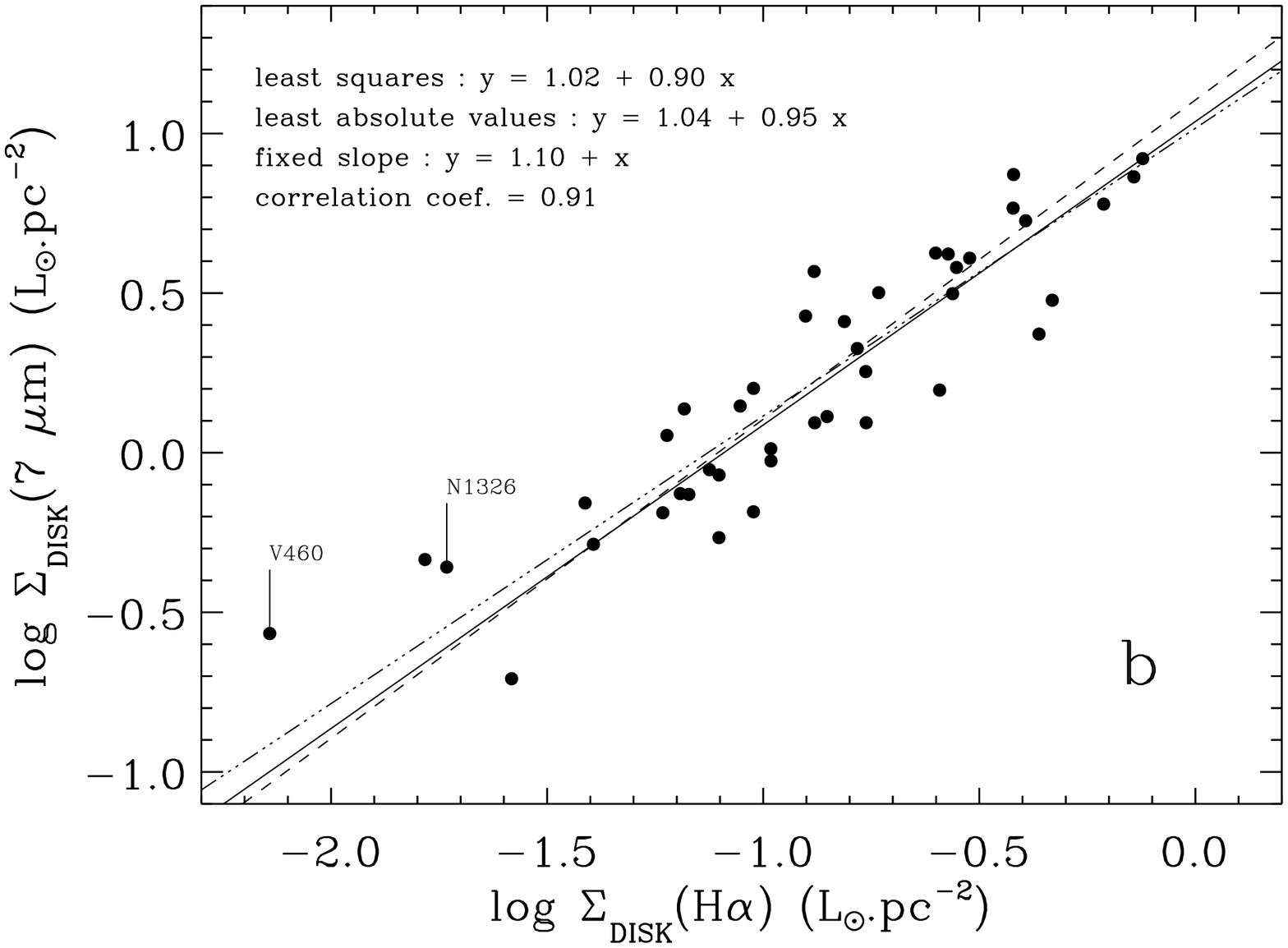}}}}
\vspace*{-0.5cm}
\centerline{\resizebox{9cm}{!}{\rotatebox{90}{\includegraphics{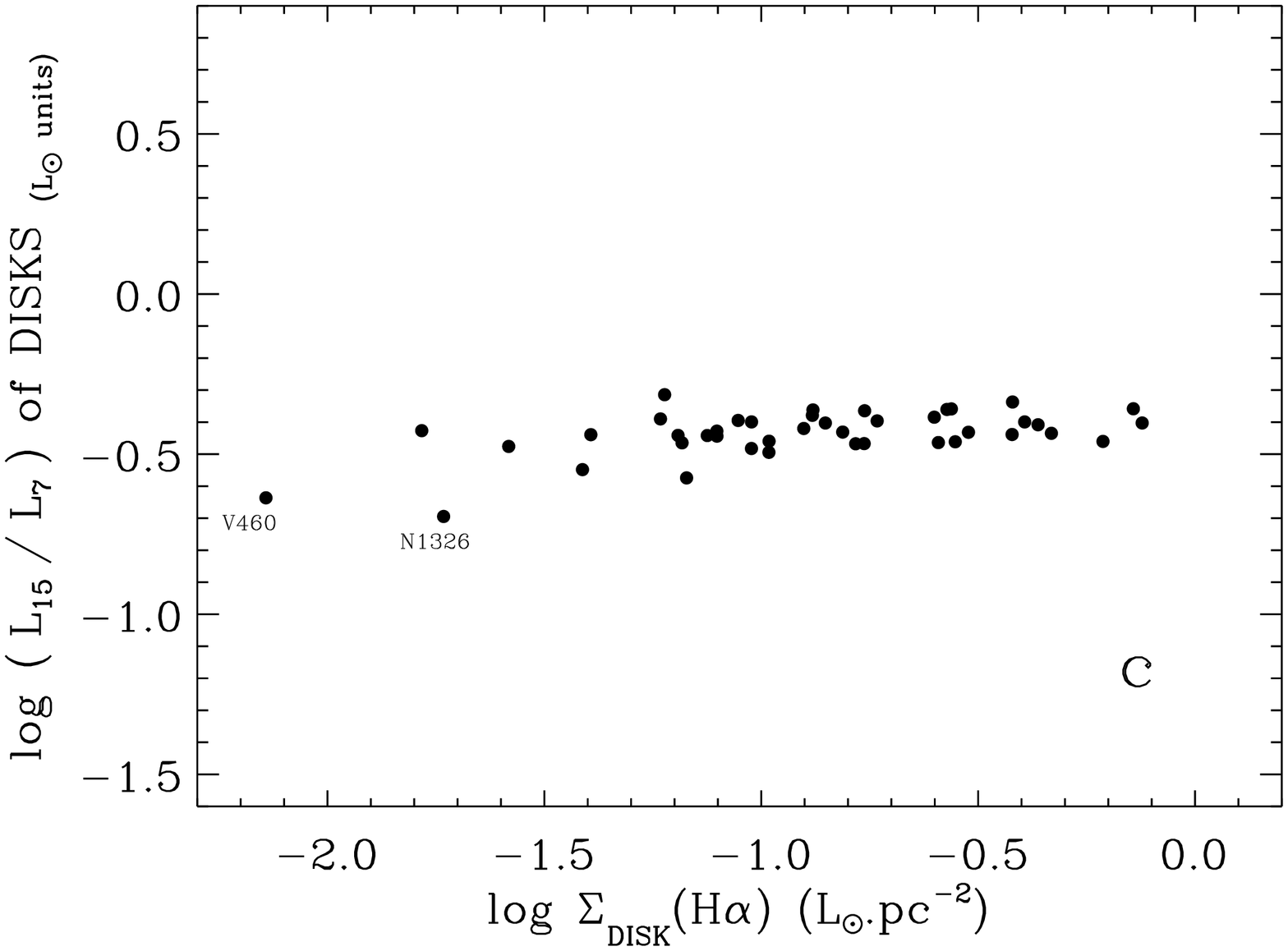}}}}
\vspace*{-0.3cm}
\caption{Relationship between size-normalized fluxes of H$\alpha$ and:
         {\bf (a)}~$F_{15}$ (12 to 18\,$\mu$m); {\bf (b)}~$F_7$ (5 to 8.5\,$\mu$m).
	 The dashed line is the linear correlation. The best least squares fit
         and least absolute deviation fit are shown as dot-dashed and solid lines.
         NGC\,4736 and NGC\,6744 were excluded because their disk was not
         completely mapped at 7 and 15\,$\mu$m. Excluding in {\bf (b)} the two
         galaxies with suspected significant contribution from stellar emission
         at 7\,$\mu$m, because of their low $F_{15}/F_7$ ratios (VCC\,460 and
         NGC\,1326), the fitted slopes become 1.00 and 1.07, instead of 0.90
         and 0.95\,. {\bf (c)} shows that $F_{15}/F_7$ is constant with a good
         approximation in disks. Flux density ratios (Jy units) can be obtained
         applying a factor 2.40 (0.38\,dex).}
\label{fig:sb_lw_ha}
\end{figure}

We applied two fitting procedures. The first one is the classical minimization
of squared distances to a line, with an equal treatment of both variables.
The second one is the minimization of absolute values instead of squares,
again bivariate, and is more robust to outlier points. Fig.~\ref{fig:sb_lw_ha}a
shows the dependence of the size-normalized 15\,$\mu$m on the size-normalized
H$\alpha$ fluxes. The best least squares fit implies that $F_{15}$ scales as
$F$(H$\alpha)^{1.01}$, with a correlation coefficient of 0.91; the 3\,$\sigma$
interval for the slope is [0.81; 1.25]. The results for 7\,$\mu$m are quite
similar to those for 15\,$\mu$m (Fig.~\ref{fig:sb_lw_ha}b), with approximately
the same power law. That the 7 and 15\,$\mu$m flux densities appear interchangeable
in Fig.~\ref{fig:sb_lw_ha} may come at first as a surprise given the already
growing amount of literature data indicating that regions of high star formation
activity appear as regions of enhanced 15\,$\mu$m emission with respect to
7\,$\mu$m \citep{Vigroux96, Sauvage96, Dale}. However it is also clear that:
(1) not all star forming regions show a 15\,$\mu$m excess \citep[see {\it e.g.}
the color data in][ and our Atlas]{Dale} and (2) the excess appears only
above a certain threshold in star-formation activity. This last point is
amply demonstrated by the so-called ISO-IRAS color-color diagram
\citep{Vigroux, Helou00} that plots the $F_{15}/F_7$
ISO color versus the $F_{60}/F_{100}$ IRAS color: for most of the $F_{60}/F_{100}$
range, the $F_{15}/F_7$ color is nearly constant and around 1, and only starts to
increase for the hottest $F_{60}/F_{100}$ colors, corresponding to starburst
and interacting galaxies. For normal spiral galaxies such as those in our sample,
the $F_{15}/F_7$ color does not significantly deviate from 1. In fact, we show
in Paper~I that changes of the global $F_{15}/F_7$ color of spirals are strictly
due to the circumnuclear regions which are not included in Fig.~\ref{fig:sb_lw_ha}.
Spiral disks exhibit a $F_{15}/F_7$ color of $0.89 \pm 0.14$ (in flux density
units, {\it i.e.} in Jy).

\begin{figure}[!b]
\centerline{\resizebox{9.5cm}{!}{\rotatebox{90}{\includegraphics{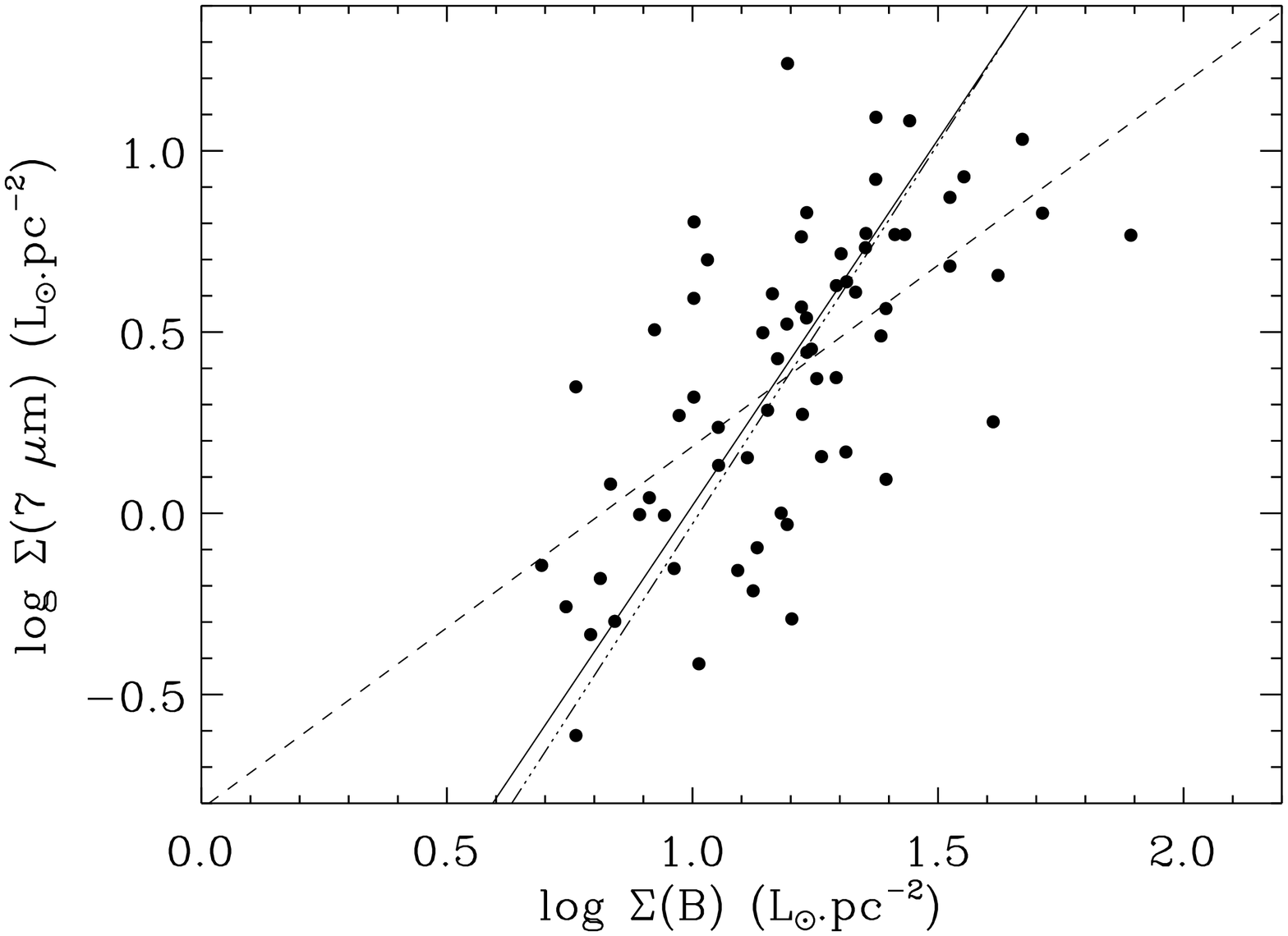}}}}
\caption{Comparison of total size-normalized fluxes in the blue band and at
         7\,$\mu$m. The dashed line is the linear correlation, and the best least
         squares fit and best least absolute deviation fit are shown as
         dot-dashed and solid lines. The fitted slopes are 2.10 and 2.02, with
         a 1\,$\sigma$ confidence interval [1.85; 2.41] and a linear correlation
         coefficient of 0.65\,. Using 15\,$\mu$m fluxes instead of 7\,$\mu$m
         fluxes or H-band fluxes for the stellar emission
         leads to very similar results. Restricting the sample to the galaxies
         present in Fig.~\ref{fig:sb_lw_ha} or to the galaxies dominated by
         disk emission at 15\,$\mu$m by more than 70\% also produces similarly
         dispersed distributions with fitted slopes always above 1.6\,.}
\label{fig:sb_lw_B}
\end{figure}

The fact that the 7 and 15\,$\mu$m fluxes show similar variations with
radiation density (in disks) is somewhat puzzling, as they were originally
thought to behave quite differently, since the main emission sites of aromatic
bands and VSGs (respectively photodissociation regions and H{\scriptsize II}
regions) are distinct. A likely explanation is that when averaged in disks, the
15\,$\mu$m emission is no longer dominated by VSGs as in H{\scriptsize II} regions
but by a part of the feature composed of the aromatic bands at 11.3 and
12.7\,$\mu$m, and thus shares a common origin with the 7\,$\mu$m emission. This
is what can be seen in our few mid-IR spectra (Paper~I). Other fainter UIBs are
additionally present in the 13-18\,$\mu$m range \citep[see for instance][]{Hony}.
A further confirmation of the common origin
of the 7 and 15\,$\mu$m fluxes in disks can be found in the fact that the
dependence of $F_{15}/F_7$ on the IRAS color $F_{25}/F_{12}$ is weak for low ratios
($F_{25}/F_{12} < 2$) and begins to strengthen only for relatively hot colors:
as the 25\,$\mu$m flux density is clearly due to VSGs, this indicates another
origin for the 15\,$\mu$m flux density in the low $F_{25}/F_{12}$ range.

Fig.~\ref{fig:sb_lw_ha} thus implies that on the scale of galactic disks,
in the mid-infrared, H{\scriptsize II} regions are seen only through their effect
of globally increasing the interstellar radiation field. Since impulsive heating
of aromatic band carriers by single photons causes the shape of the spectrum
to be very insensitive to the radiation intensity \citep{Boulanger}, this
explains the constant $F_{15}/F_7$ ratio observed in our sample. It results from
this that even over the large range in H{\scriptsize II} region sizes and densities
seen along the Hubble sequence, what we observe in disks is mainly the emission
from photodissociation regions, and the filling factor by H{\scriptsize II}
regions is always comparatively small.

Although mid-IR emission can in principle originate from regions where old
stars dominate, when integrated throughout spiral disks, the emission in both
bandpasses traces young stars and the heating provided by a more evolved
population appears negligible. Otherwise, the $F_{15}$/H$\alpha$ ratio would
increase with decreasing SFR surface density, as the importance of the
diffuse interstellar medium relative to star forming regions is then higher;
this is not observed, and neither is a variation of $F_{15}$/H$\alpha$ with
Hubble type.

An alternative cause of the observed correlation could be a much more
indirect link between star formation and dust emission, both of them being
causally associated with the molecular gas phase (dust is mixed with gas,
and stars form out of molecular clouds), and these two links then simulate a
direct connection between the presence of massive stars and dust excitation.
We have no means of deciding which scenario is the more likely, and
they would observationally be extremely difficult to test, in particular
because molecular gas mass estimates are not accurate enough. Nevertheless,
the presence of molecular gas is a prerequisite but certainly not a sufficient
condition for star formation, so that the link between the two is not more
direct than the link between young stars and dust emission.

Assuming a purely linear correlation between UIB and H$\alpha$ emission
(as indicated in Fig.~\ref{fig:sb_lw_ha} by the dashed line) and using
the H$\alpha$-SFR calibration of \citet{Kennicutt98a} leads to the following
scalings:
\begin{eqnarray}
{\rm SFR}~({\rm M}_{\odot}\,{\rm yr}^{-1})&=&6.5~10^{-9}~L{\rm (15~\mu{\rm m}_{~UIB})~(L_{\odot bol}}) \nonumber \\
                                          &=&2.4~10^{-9}~L{\rm (7~\mu{\rm m}_{~UIB})~(L_{\odot bol}})
\label{eq:calib}
\end{eqnarray}
with $L_{\odot {\rm bol}} = 3.827~10^{26}$\,W (this assumes solar abundances,
and a Salpeter IMF ranging from 0.1 to 100\,M$_{\odot}$). Flux densities at 7 and
15\,$\mu$m were converted into luminosities using bandpasses of 16.18\,THz and
6.75\,THz respectively. These formulas are applicable only when the mid-infrared
emission is dominated by UIBs, with a negligible VSG continuum, which is the
case in disks of galaxies, but is not always verified at 15\,$\mu$m (in galactic
central regions, for instance).

In our sample, size-normalized SFRs in disks (we mean here physical regions
defined in Paper~I) range between about 0.3 and
20\,M$_{\odot}$\,kyr$^{-1}$\,kpc$^{-2}$, from VCC\,1043 to NGC\,5236, and SFRs
of disks between about 0.12 and 8\,M$_{\odot}$\,yr$^{-1}$.

We emphasize that the relationship found between dust emission and the
H$\alpha$ recombination line is unique and is not reproduced if the H$\alpha$
emission is replaced by an observable tracing more evolved stellar populations,
such as the blue luminosity, coming from the RC3 (Fig.~\ref{fig:sb_lw_B}).

\section{Discussion}
\label{sec:dis}

\subsection{Integrated versus local quantities}

The above calibration, derived from integrated disk fluxes, should not be
applied locally in small regions of disks, for several reasons:
\begin{list}{--}{}
\item In the diffuse interstellar medium, far away from any star-forming
      site, a substantial part of UIB carrier heating may be provided by
      optical photons, since it has been demonstrated, for instance in
      reflection nebul\ae, that this type of heating can be efficient
      \citep{Uchidaa}. Looking at quantities integrated over large
      spatial scales, we average in our beam all the stellar populations
      contributing to the excitation of dust. Since dust heating by
      ionizing photons is much more efficient than by near-ultraviolet or
      optical photons, a small population of massive stars mixed with a
      large population of low-mass stars can still dominate dust heating as
      soon as it exceeds a certain threshold, still to be determined. It
      is clear that this could no longer be the case in selected regions of
      galaxies.
\item Close to H{\scriptsize II} regions and in some giant star formation complexes,
      the $F_{15}/F_7$ color is observed to rise above the mean value in disks (see
      for instance NGC\,5457 in the Atlas), revealing a thermodynamical state
      different from that in disks for the species emitting at these two
      wavelengths. This 15\,$\mu$m excess clearly breaks down the symmetry of the
      15 and 7\,$\mu$m fluxes, and then most likely the validity of our calibration.
\item A significant fraction of ionizing photons is able to escape H{\scriptsize II}
      regions, and possibly to propagate to very large distances. In such a
      configuration, a physical link between young stars and dust heating can be
      established only using integrated fluxes.
\end{list}

To support the last caveat, we note that \citet{Beckman} propose, from
their study of H$\alpha$ luminosity functions of H{\scriptsize II} regions, that
the estimated ionizing luminosity able to escape the most luminous H{\scriptsize II}
regions is more than sufficient to ionize the warm diffuse interstellar medium
\citep[see also][ and references therein]{Oey}; it is hence also able to heat dust
to comparable distances. If the idea that the diffuse ionized gas is excited by
the photons escaping from H{\scriptsize II} regions is confirmed, then the high
H$\alpha$ luminosity fractions attributable to this diffuse medium
\citep[30--50\% according to {\it e.g.} ][]{Ferguson} imply that a correct
assessment of star formation activity cannot be made if restricted to limited
regions, neither using recombination lines nor using the infrared emission.

\citet{Vogler} have looked at the relationship between H$\alpha$ and
7\,$\mu$m fluxes locally inside the disk of NGC\,5236, using resolution
elements of $12\arcsec$ ($\approx 300$\,pc at our adopted distance). They
find a very large dispersion, mostly at low flux levels, so that the correlation
that we report here breaks down at kpc-scales. It is however difficult to
disentangle in the observed scatter the role played by variable extinction
in the H$\alpha$ line (spatially and also as a function of brightness) from
a physical decorrelation.

\subsection{Mid-IR versus far-IR emission as tracers of star formation}
\label{sec:fir}

In Sect.~\ref{sec:sfr}, restricting ourselves to spiral disks, we have found
that mid-IR luminosities can be used to trace the level of star formation. As
mentioned in Sect.~\ref{sec:intro}, numerous attempts to use the far-IR
luminosity as a tracer of star formation have already been presented
\citep[see {\it e.g.} ][ and references therein]{Devereux, Kennicutt98a}. It
is thus worthwhile to examine whether the tracer we propose here presents some
significant advantage over the far-IR one. The main problem with the far-IR
luminosity is that the heating radiation required for grains to emit in that
wavelength range can easily be provided by relatively old stars of
10$^{8}$~yr or more \citep{Buat} and thus the information derived on the
SFR is different from that obtained from H$\alpha$. In other terms, when plotting
far-IR versus H$\alpha$ data, there is a hidden variable which corresponds to the
amount of energy provided by non-ionizing stars. It is this hidden variable
which has previously been deemed responsible of the significant
non-linearity of the far-IR--H$\alpha$ correlation \citep[{\it e.g.} ][]{Lonsdale}.

\begin{figure}[!t]
\centerline{\resizebox{9.5cm}{!}{\rotatebox{90}{\includegraphics{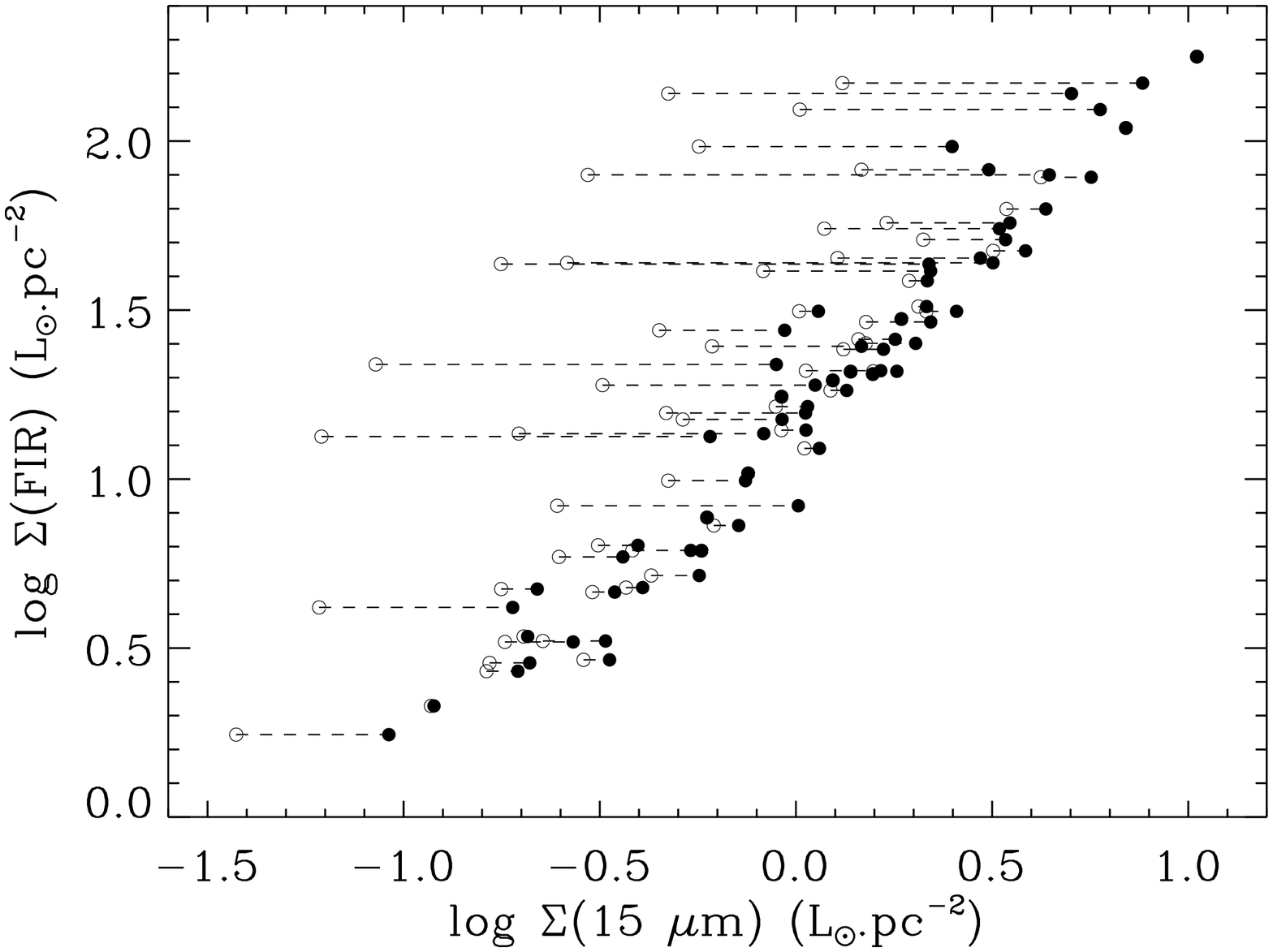}}}}
\caption{Proportionality between far-infrared and mid-infrared size-normalized
         fluxes. Far-IR fluxes are a combination of 60 and 100\,$\mu$m IRAS fluxes
         as defined in \citet{Helou88}. Filled circles represent total fluxes
         and open circles show the effect of taking into account only bare disks,
         as defined in Paper~I (this was possible only at 15\,$\mu$m, since galaxies
         are not resolved by IRAS). The best least-squares fit gives a slope of
         1.08 ($^{+0.11}_{-0.10}$ at a 3\,$\sigma$ confidence level) with a
         correlation coefficient of 0.97\,. The relationship with 7\,$\mu$m fluxes
         is similar, but more dispersed and with a slightly higher slope
         ($1.20^{+0.18}_{-0.15}$ at a 3\,$\sigma$ confidence level).}
\label{fig:fir_LW3}
\end{figure}

Fig.~\ref{fig:fir_LW3} shows that the two integrated far-IR and 15\,$\mu$m
size-normalized fluxes are tightly and linearly correlated. Consequently,
there cannot be major differences in the heating source for the small
grains responsible for the 15\,$\mu$m emission and the large grains emitting
in the far-infrared. If we restrict ourselves to galaxies with small
circumnuclear contribution (in Fig.~\ref{fig:fir_LW3}, those for which
the two symbols used are close to each other), this suggests that in galaxies
which are dominated by their disk, the far-infrared emission is an equivalently
good star formation tracer as the 15 and 7\,$\mu$m emission.

\begin{figure}[!t]
\centerline{\resizebox{9.5cm}{!}{\rotatebox{90}{\includegraphics{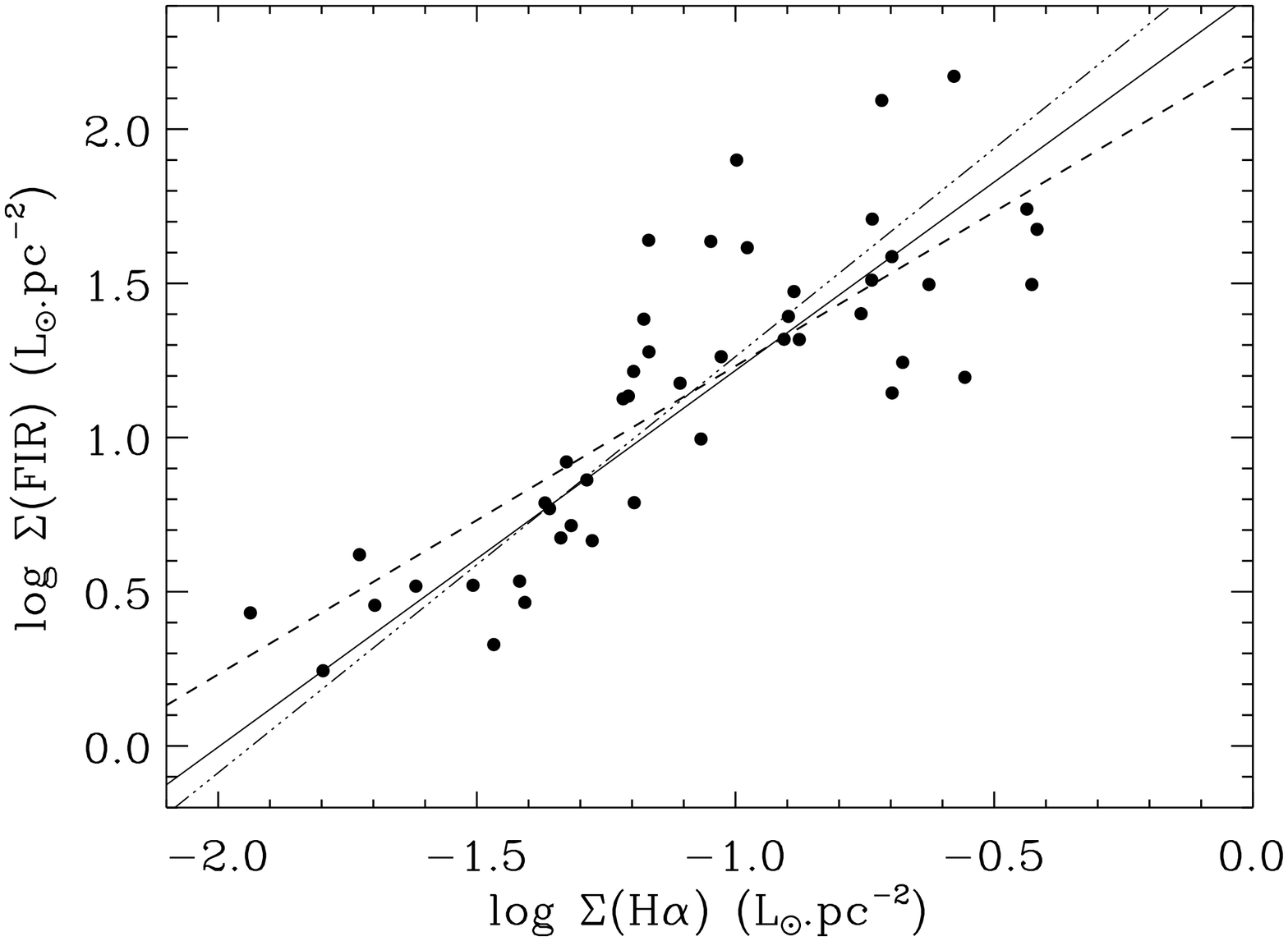}}}}
\caption{Relationship between total far-IR and H$\alpha$ size-normalized fluxes.
         The dashed line represents the linear correlation, the dot-dashed line
         the least squares fits, and the solid line the least absolute
         deviation fit, with respective slopes $1.35^{+0.50}_{-0.34}$
         (3\,$\sigma$ interval) and 1.22\,.}
\label{fig:firhalpha}
\end{figure}

\begin{figure}[!t]
\centerline{\resizebox{9.5cm}{!}{\rotatebox{90}{\includegraphics{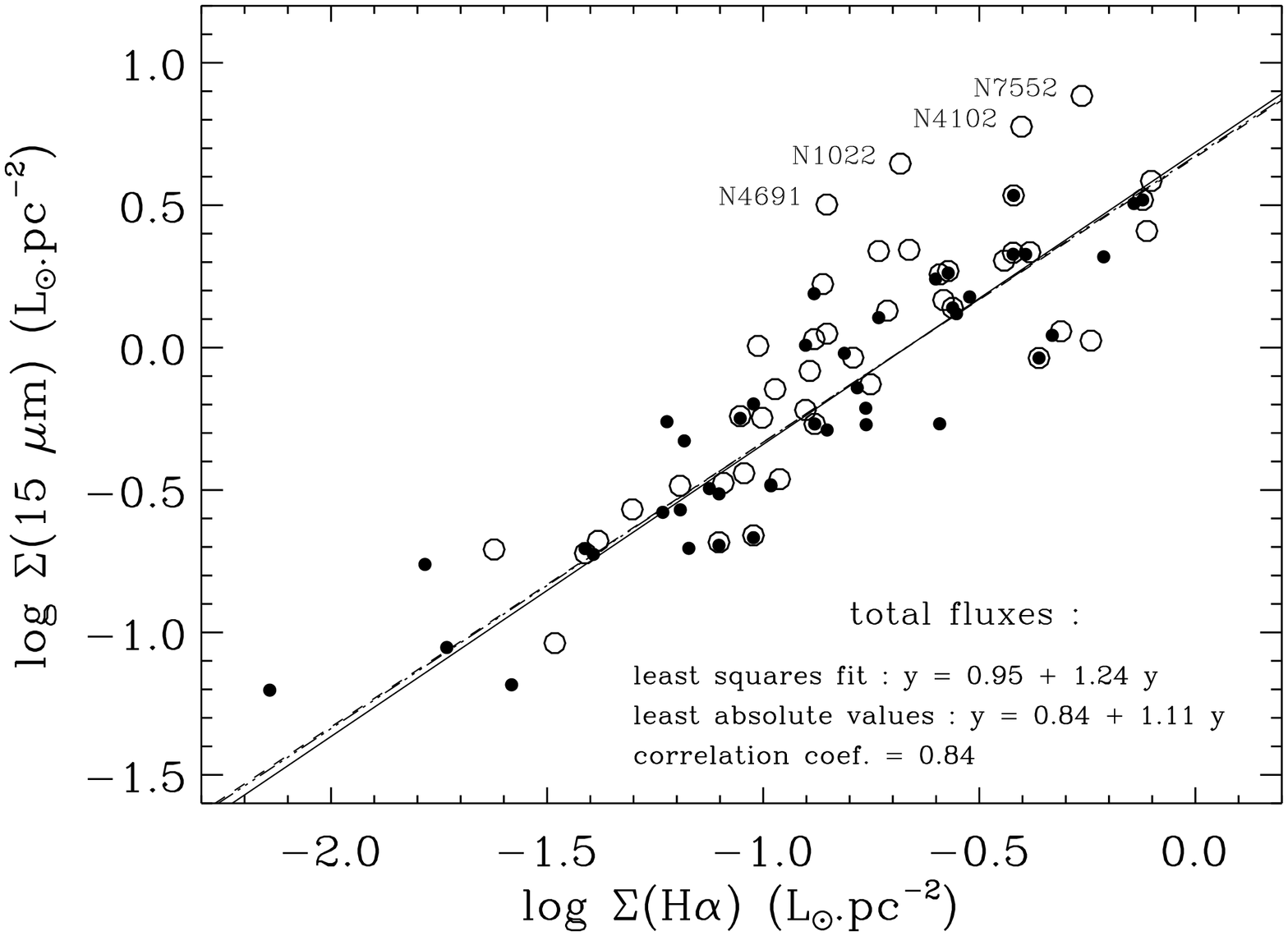}}}}
\caption{Same as Fig.~\ref{fig:sb_lw_ha}, with {\em total} size-normalized
         15\,$\mu$m and H$\alpha$ fluxes superimposed, shown as empty circles. The
         solid and dot-dashed lines correspond to the fits on disk fluxes of
         Fig.~\ref{fig:sb_lw_ha}.
         The names of the most strongly deviating galaxies and the numerical
         results of the fits on total fluxes are indicated. At 7\,$\mu$m, only
         NGC\,1022 and 4691 tend to stray from the relationship found in disks.}
\label{fig:sb_lw_ha_tot}
\end{figure}

In Fig.~\ref{fig:firhalpha}, we plot the correlation between the
size-normalized {\em total} far-IR and H$\alpha$ fluxes for galaxies in our sample.
Given the very poor spatial resolution of current far-IR instruments, it is
impossible to separate nuclear and disk contributions. We see here that
the correlation is worse than that obtained in Fig.~\ref{fig:sb_lw_ha} and in
particular that the well-known non-linearity of the correlation is found in
our sample too: a least-square fit to our data gives a slope
of 1.35\,.\footnote{Using a large sample drawn from the H$\alpha$ catalogues
mentioned in Sect.~\ref{sec:sample} and applying a robust estimation method to
reduce the effects of outlying points, since a significant scatter is
present, we obtain $\Sigma(F_{\rm FIR}) \propto \Sigma$(H$\alpha)^{1.43}$.}.
The results obtained are highly dependent on the adopted references for IRAS
fluxes among the various catalogs
\citep[][; PSC]{Moshir, Rice, Soifer, Helou88, Sanders, Rush, Thuan}, which may
partially account for the dispersion.
However, we have tried to identify the most reliable fluxes by optimizing the
agreement between our 7 and 15\,$\mu$m fluxes and IRAS 12\,$\mu$m fluxes, whose
bandpass overlaps with both ISOCAM bandpasses.

Looking for the origin of this non-linearity, we find that it is mainly
introduced in our sample by prominent circumnuclear regions. Indeed, when
the fraction of the 15\,$\mu$m flux contributed by central regions increases
(from nearly 0 to nearly 1), the mean $F_{\rm FIR}/F({\rm H}\alpha)$ ratio rises
from $\approx 130$ to $\approx 480$. The increase of $F_{\rm FIR}/F({\rm H}\alpha)$
is also connected with the appearance of high $F_{15}/F_7$ colors, which have been
shown in Paper~I to originate in central regions of galaxies whose total
mid-infrared emission is dominated by a central starburst.
We can further check this finding by estimating the fraction of far-IR emission
arising from the disk alone and producing an analog of Fig.~\ref{fig:sb_lw_ha}
for the far-IR. Since no spatially-resolved measurements are available, we
have to rely on disk flux fractions measured at 15\,$\mu$m and assume that they
can account for the true fractions in the far-IR. Given that total 15\,$\mu$m and
far-IR fluxes are tightly correlated, this assumption is sensible.
This empirical correction succeeds in linearizing the far-IR--H$\alpha$ relationship
and reducing the scatter: the least squares and least absolute deviation fits
give a slope of respectively $1.01^{+0.25}_{-0.20}$ and 1.07\,.
This demonstrates that it is the circumnuclear contribution which creates the
non-linearities observed in previous attempts to correlate the far-infrared and
H$\alpha$ fluxes.

To further support this view, we have looked at correlations between
{\em total} mid-infrared and H$\alpha$ fluxes, adding the two galaxies whose
mid-infrared and H$\alpha$ emission is highly concentrated, NGC\,1022 and 4691.
Fig.~\ref{fig:sb_lw_ha_tot} shows the result at 15\,$\mu$m, compared
with what we have previously obtained in disks. Owing to the fact that H$\alpha$
data were found for only few centrally dominated galaxies with a 15\,$\mu$m excess,
the relationship for total fluxes is not much different from that for disk
fluxes. However, the fits indicate that the dispersion is already higher and
that we introduced a non-linearity. As the extinction in circumnuclear regions
is expected to be higher than in the average disk, the interpretation of the
cause for the non-linearity is not straightforward. We postpone the detailed
study of mid-infrared emission and other star formation estimates in
circumnuclear and starburst regions (F\"orster Schreiber \& Roussel, in
preparation), and outline that the relationship found here is strictly valid
only in disks, and cannot be extrapolated easily.

\subsection{Applicability at high redshift}

In cosmological surveys with the HST, SFRs are measured with fluxes that
correspond to ultraviolet wavelengths in the rest-frame of distant galaxies.
For this reason, we have tried to compare the SFRs deduced from mid-infrared
fluxes using Eq.~\ref{eq:calib} with the SFRs derived from ultraviolet fluxes
at 1650\,\AA\ using the calibration given in \citet{Kennicutt98a}. The data come
from \citet{Rifatto}, \citet{Deharveng} and \citet{Bell} (whose filters are
not centered exactly at 1650\,\AA\ but at 1615 and 1521\,\AA), and the sample was
limited to galaxies either with a disk flux fraction above 0.7 or with
$F_{15}/F_7 \leq 1.2$ (respectively 20 and 29 objets), both criteria leading to
the same conclusion. UV fluxes have only been corrected for Galactic extinction
in the same way as H$\alpha$, and the internal extinction has been estimated by the
requirement that the 7 or 15\,$\mu$m and ultraviolet SFR values agree for
each spiral. Resulting UV absorptions range between 0.2 and 2.6\,mag, and
their median value amounts to 1.4--1.7\,mag. We recall that our calibration
assumes an H$\alpha$ absorption of 1.1\,mag, and that changing $A$(H$\alpha$) would
change $A_{1650}$ by the same amount.

We can compare these numbers with those obtained in the high-redshift sample
of \citet{Flores} ($0.2 < z < 1.1$), who have performed a similar comparison
between star formation rates derived from ultraviolet and infrared observations.
They have extrapolated ultraviolet fluxes at 2800\,\AA\ from spectral energy
distributions above 4350\,\AA, using a grid of spectro-photometric evolutionary
models, and have also estimated total infrared fluxes from 8 to 1000\,$\mu$m from
mid-infrared and radio fluxes, fitted by a set of templates. Assuming the same
IMF as in the present work and the SFR calibrations of \citet{Kennicutt98a}
both in terms of UV and IR fluxes, \citet{Flores} derive extinctions in the
range $0.5 \leq A_{2800} \leq 2.2$\,mag. This translates into $0.6 < A_{1650} < 2.8$,
using the extinction curves of \citet{Cardelli}. Absorption estimates in our
sample are thus consistent, well within the uncertainties, with those in the
sample of \citet{Flores}. Hence, although absorption estimates can be
flawed by metallicity effects in young galaxies -- if they deplete preferentially
small carbonaceous grains emitting in the mid-infrared -- there is no hint of a
significant increase of optical depths with redshift, which implies that the
calibration we give likely remains valid in more distant galaxies.

\section{Summary and conclusions}

We have seen that mid-infrared fluxes can be considered reliable tracers of
star formation in relatively quiescent environments such as spiral galactic disks.
There, mid-infrared spectra between 5 and 18\,$\mu$m are dominated by aromatic
bands, which are much more tightly linked with the radiation from young stars
than with the blue-band radiation, to which stars of all masses and ages
contribute.

The calibration in terms of SFR that we propose depends of course on the
adopted IMF, but most sensitively on the extinction correction applied to H$\alpha$
fluxes. The dispersion in H$\alpha$ fluxes with respect to the linear correlation is
0.19 dex in Fig.~\ref{fig:sb_lw_ha}, much lower than the extinction correction
of 0.44 dex. We find a bivariate dispersion around the linear correlation of a
factor of 1.37 and a one-dimensional dispersion of a factor of 1.56, both at 7 and
15\,$\mu$m. Despite this large scatter, it is meaningful to apply our calibration
to large samples, because the size-normalized SFR shows a much larger range
(it varies by more than a factor of 50 across the present sample).

Nevertheless, one must be aware of several limitations: \\
-- Because dust grains can be heated by the radiation from H{\scriptsize II}
   regions at large distances from them, and because dust, in some diffuse regions,
   can be predominantly heated by old stellar populations, the relationship between
   mid-infrared fluxes and star formation rates is certainly much more complex
   locally than when considering integrated fluxes. \\
-- For distant objects detected in surveys, the only available information
   consists of fluxes integrated over the whole galaxy, which can be dominated
   by the central regions. In this case, the link between star formation and
   mid-infrared emission can also be completely different. Indeed, we show that
   considering global fluxes, in the far-infrared as well as at 15\,$\mu$m,
   introduces a non-linearity in the correlation with H$\alpha$ fluxes. This may
   be due to a combination of several effects: a thermodynamical state of dust
   grains in central regions different from that in disks; a greater extinction
   affecting the H$\alpha$ line; a dominant contribution from post-starburst
   populations of the bulge to dust heating in the absence of significant star
   formation. \\
-- Our sample comprises only spiral galaxies whose metallicity is thought to
   be near solar, while metal-deficiency (mostly seen in blue compact and
   dwarf irregular galaxies) can alter the dust composition and is likely to
   deplete carbonaceous grains, which tends to lower the mid-infrared emission
   for a given radiation field \citep{Sauvage90, Boselli}.
   The aromatic bands are often absent or very weak in dwarf galaxies, due to
   the above effect combined with their destruction by the far-ultraviolet
   radiation, very pervasive in low-metallicity environments \citep{Madden}. \\
-- In extremely active environments with normal metallicity, the aromatic band
   carriers can also be destroyed or experience chemical transformations, but
   quantitative estimations of these effects are yet unavailable.

However, this relationship between mid-IR fluxes and SFRs that strictly holds
only in normal disks can be useful to interpret surveys made in the two filters
LW3 (15\,$\mu$m) and LW2 (7\,$\mu$m). For galaxies at high redshifts
($z \simeq 1.2$), the LW2 rest frame emission is shifted to the LW3 bandpass.
Hence, the calibration given here must provide a lower limit for the true SFR
since, for galaxies with greater star formation activity than in the present
sample, the energy redistribution favors the LW3 band, as more energy is reradiated
by VSGs. Indeed, \citet{Boulanger3} have presented observations in resolved Galactic
regions (a diffuse cloud and four photodissociation regions) which show that
the emission in UIBs tends to rise with the ultraviolet energy density, linearly
at low energy densities and more slowly at higher values. The threshold for this
transition, above $10^3$ times the energy density in the solar neighborhood,
is uncertain due to dilution effects. However, the same type of behavior must
hold in integrated galaxies.

Dealing with integrated fluxes, a simple validity criterion of the formula
presented here would be a rest-frame color $F_{15}/F_7 \simeq 1$, {\it i.e.} the
contribution from the central concentration to the total emission should be low
or should arise from heating by a disk-like (non-starburst) stellar population.

\begin{acknowledgements}
We warmly thank Antonio Garc\'{\i}a-Barreto, Thaisa Storchi-Bergmann,
Magnus Naslund, Carlos Feinstein, Michael Regan, Stuart Ryder and
Fran\c{c}ois Viallefond for freely providing their H$\alpha$ maps.
Our second referee, Martin Haas, deserves our gratitude for his rapid answer
and useful suggestions. We wish to thank Suzanne Madden for taking part
in the improvement of the manuscript and Herv\'e Aussel for a discussion
about high redshift studies of star formation.
\vspace*{1ex} \\
The ISOCAM data presented in this paper were analyzed using and adapting
the CIA package, a joint development by the ESA Astrophysics Division and the
ISOCAM Consortium (led by the PI C. Cesarsky, Direction des Sciences de la
Mati\`ere, C.E.A., France).
\end{acknowledgements}

\bibliographystyle{apj}

\clearpage

\begin{table*}[!t]
\caption[]{Photometric data. Galaxies are named according to the VCC catalog
           for the {\it Virgo} program and according to the NGC catalog for
           other programs.}
\label{tab:tableau}
\begin{minipage}[t]{15cm}
\begin{flushleft}
\begin{tabular}{|l@{\hspace*{-2ex}}crcccrrr@{, \hspace*{-1mm}}r|}
\hline
\noalign{\smallskip}
name & (H$\alpha$~+~[N{\scriptsize II}])$_{\rm TOT}$\footnote{corrected for
   Galactic extinction using the RC3 blue absorptions and the extinction curve
   of \citet{Cardelli}. Total pure H$\alpha$ fluxes, available for N1530, N5383,
   N6744 and N6946, were made homogeneous with (H$\alpha$~+~[N{\scriptsize II}])
   fluxes by applying a factor $4/3$.} &
$D_C$\footnote{diameter aperture or slit dimensions of the measurement used to
   remove a central H$\alpha$ contribution.} &
(H$\alpha$~+~[N{\scriptsize II}])$_{\rm CEN}$ $^a$ &
$f_{C\, 15}$\footnote{approximate fractions of central mid-infrared fluxes inside
   the same aperture as used for H$\alpha$ (corrected for dilution effects).} &
$f_{C\, 7}$ $^c$ & $F_{15\: \rm DISK}$\footnote{mid-infrared fluxes after removal
   of the central contribution matched to the H$\alpha$ aperture.} &
$F_{7\: \rm DISK}$ $^d$ & \multicolumn{2}{c|}{refs$^e$} \\
~ & (log W\,m$^{-2}$) & (arcsec) & (log W\,m$^{-2}$) & ~ & ~ &
\multicolumn{2}{c}{(mJy)} & \multicolumn{2}{c|}{~} \\
\noalign{\smallskip}
\hline
\noalign{\smallskip}
 N337                  &  -14.38 &     $4. \times 4.$ & -15.69 &  0.03       &  0.03       &   288.27 &   327.34 & 11a & 21 \\
N1022                  &  -14.91 &                  ~ &      ~ & $\approx 1$ & $\approx 1$ &        ~ &	       ~ &  8  & .. \\
N1097                  &  -13.94 &                45. & -14.27 &  0.76       &  0.60       &   540.84 &   846.37 & 20  & 20 \\
N1365                  &  -13.74 &                40. & -14.32 &  0.71       &  0.54       &  1306.65 &  1701.55 & 12  & 12 \\
N1433                  &  -14.67 &                26. & -15.39 &  0.31       &  0.26       &   246.95 &   283.21 &  2  &  2 \\
N1530~$^{(-)}$         &  -14.46 &                28. & -15.32 &  0.44       &  0.38       &   337.69 &   356.16 & 14  & 14 \\
N1672                  &  -13.93 &                32. & -14.40 &  0.58       &  0.48       &   842.58 &  1030.42 & 20  & 20 \\
N4027                  &  -14.41 &     $4. \times 4.$ & -16.21 &  0.01       &  0.01       &   668.50 &   765.68 & 11a & 21 \\
N4535                  &  -14.26 &      $\approx 10.$ & -15.22 &  0.11       &  0.08       &  1000.65 &  1047.02 & 24  & 13 \\
N4691                  &  -14.94 &                  ~ &      ~ & $\approx 1$ & $\approx 1$ &        ~ &        ~ &  6  & .. \\
N4736                  &  -13.27 &      $\approx 10.$ & -15.65 &  0.06       &  0.05       &  3970.51 &  3702.07 & 24  & 13 \\
N5194                  &  -13.33 &                89. & -14.10 &  0.25       &  0.22       &  5969.43 &  6728.27 & 24  &  7 \\
N5236                  &  -12.87 &                33. & -13.97 &  0.17       &  0.14       & 16737.02 & 15933.28 & 11b & 16 \\
N5383~$^{(-)}$         &  -14.23 &               22.3 & -14.49 &  0.49       &  0.41       &   169.72 &  205.100 & 18  & 18 \\
N5457                  &  -13.17 &                35. & -15.18 &  0.02       &  0.02       &  5295.81 &  5920.03 & 11b & 22 \\
N6744~$^{(-)}$         &  -13.54 &     $4. \times 4.$ & -16.95 &  0.01       &  0.01       &  1494.72 &  2414.59 & 17  & 21 \\
N7552                  &  -14.19 &               21.3 & -14.50 &  0.83       &  0.69       &   475.78 &   574.27 &  5  &  5 \\
\multicolumn{10}{|c|}{~} \\
  V66 (N4178)          &  -14.59 &     $2. \times 4.$ & -17.22 &  0.02       &  0.01       &   178.19 &   225.74 & 24  &  9 \\
  V92 (N4192)          &  -14.39 &       $\approx 5.$ & -15.61 &  0.06       &  0.03       &   591.66 &   872.36 & 24  & 13 \\
 V460 (N4293)          &  -14.90 &      $\approx 12.$ & -14.99 &  0.67       &  0.29       &    62.34 &   112.59 & 24  & 13 \\
 V692 (N4351)          &  -15.43 &                 8. & -16.18 &  0.12       &  0.12       &    40.28 &    46.48 & 24  & 19 \\
 V836 (N4388)          &  -14.50 &                12. & -14.97 &  0.73       &  0.38       &   267.86 &   308.83 & 11b &  4 \\
 V857 (N4394)          &  -14.96 &      $\approx 10.$ & -16.56 &  0.09       &  0.09       &   127.13 &   147.36 & 24  & 13 \\
 V912 (N4413)          &  -15.13 &                  ~ &      ~ &     ~       &     ~       &    92.98 &    89.25 & 24  & .. \\
V1043 (N4438)          &  -14.61 &                 8. & -15.28 &  0.29       &  0.20       &   149.30 &   186.32 & 24  & 19 \\
V1110 (N4450)          &  -15.17 &      $\approx 10.$ & -15.68 &  0.11       &  0.09       &   150.36 &   167.55 & 24  & 13 \\
V1379 (N4498)          &  -15.01 &                 8. & -16.53 &  0.05       &  0.05       &    89.63 &   107.73 & 24  & 19 \\
V1673 (N4567)~$^{(+)}$ &  -14.93 &                 8. & -16.20 &  0.05       &  0.04       &   279.19 &   306.28 & 24  & 19 \\
V1676 (N4568)~$^{(+)}$ &  -14.53 &                 8. & -15.88 &  0.08       &  0.06       &  1016.44 &  1014.13 & 24  & 19 \\
V1690 (N4569)          &  -14.22 &                 8. & -15.21 &  0.19       &  0.08       &   758.34 &   776.63 & 24  & 19 \\
V1727 (N4579)          &  -14.45 &      $\approx 10.$ & -14.93 &  0.17       &  0.07       &   513.84 &   624.93 & 24  & 13 \\
V1972 (N4647)          &  -14.66 &                8.1 & -16.09 &  0.04       &  0.03       &   454.66 &   459.85 & 24  & 10 \\
V1987 (N4654)          &  -14.32 &      $\approx 10.$ & -15.64 &  0.05       &  0.05       &   962.89 &  1000.25 & 15  & 13 \\
V2058 (N4689)          &  -14.78 &                 8. & -16.83 &  0.02       &  0.02       &   323.91 &   335.34 & 11a & 19 \\
\multicolumn{10}{|c|}{~} \\
 N986                  &  -14.47 &               17.8 & -15.00 &  0.57       &  0.36       &   454.91 &   512.01 &  8  &  8 \\
N1326                  &  -14.71 &                26. & -14.78 &  0.85       &  0.69       &    42.17 &    87.09 &  2  &  2 \\
N1385                  &  -14.31 &     $4. \times 4.$ & -15.93 &  0.02       &  0.01       &   768.69 &   804.42 & 11a & 21 \\
N3885                  &  -14.96 &               14.8 & -15.13 &  0.75       &  0.75       &    99.60 &    85.69 &  8  &  8 \\
N4041                  &  -14.25 &                  ~ &      ~ &     ~       &     ~       &   751.86 &   792.81 & 24  & .. \\
N4102                  &  -14.43 & $34.4 \times 34.4$ & -14.66 &  0.88       &  0.69       &   207.48 &   254.07 &  1  &  1 \\
N4519                  &  -14.74 &                 8. & -16.31 &  0.28       &  0.08       &   168.59 &   162.81 & 24  & 19 \\
N4713                  &  -14.49 &                  ~ &      ~ &     ~       &     ~       &   209.41 &   223.64 & 15  & .. \\
N5962                  &  -14.62 &     $4. \times 4.$ & -16.50 &  0.02       &  0.01       &   500.70 &   479.09 & 11a & 21 \\
N6753                  &  -14.63 &                  ~ &      ~ &     ~       &     ~       &   646.77 &   586.44 &  2  & .. \\
N6946~$^{(-)}$         &  -12.98 &                45. & -13.70 &  0.19       &  0.11       &  8640.46 & 10401.74 & 23  &  3 \\
N7218                  &  -14.75 &                  ~ &      ~ &     ~       &     ~       &   273.49 &   260.59 & 11a & .. \\
\noalign{\smallskip}
\hline
\end{tabular}
\end{flushleft}
\end{minipage}
\end{table*}

\clearpage

\begin{minipage}[t]{15cm}
~~$^e$ {\small Reference codes for the total H$\alpha$ flux and the central
H$\alpha$ flux. \\
(1)~\citet{Armus}; (2)~\citet{Crocker}; (3)~\citet{Engargiola};
(4)~\citet{Falcke}; (5)~map provided by C. Feinstein; (6)~ map provided by
J.A. Garc\'{\i}a-Barreto; (7)~map taken from the electronic edition of
\citet{Greenawalt}; (8)~\citet{Hameed}; (9)~\citet{Ho}; (10)~\citet{Keel};
(11a)~\citet{Kennicutt83b}; (11b)~\citet{Kennicutt94}; (12)~map provided by
M. Naslund; (13)~\citet{Pogge}; (14)~map provided by M. Regan;
(15)~\citet{Romanishin}; (16)~map provided by S.D. Ryder; (17)~\citet{Ryder};
(18)~\citet{Sheth}; (19)~\citet{Stauffer}; (20)~map provided by
T. Storchi-Bergmann; (21)~\citet{Veron}; (22)~map provided by F. Viallefond;
(23)~\citet{Wang}; (24)~\citet{Young}. \\
The correction for stellar absorption of \citet{Ho} was cancelled, using
their data. \\
\hspace*{1ex} $^{(+)}$ close system: we used the total H$\alpha$ flux listed
by \citet{Young} weighted by the two contributions inside a smaller aperture
derived from \citet{Kennicutt87}. \\
\hspace*{1ex} $^{(-)}$ (H$\alpha$~+~[N{\scriptsize II}]) flux estimated from
a pure H$\alpha$ flux.}
\end{minipage}

\end{document}